\documentclass[11pt]{article}
\newcommand{\bc}{\begin{center}}
\newcommand{\ec}{\end{center}}
\newcommand{\be}{\begin{equation}}
\newcommand{\ee}{\end{equation}}
\newcommand{\ba}{\begin{array}}
\newcommand{\ea}{\end{array}}
\newcommand{\bea}{\begin{eqnarray}}
\newcommand{\eea}{\end{eqnarray}}
\newcommand{\edc}{\end{document}}

\usepackage{graphics}
\usepackage{amsmath,amsxtra,amssymb,latexsym, amscd,color}
\usepackage[mathscr]{eucal}
\usepackage{graphicx}
\begin{document}
\thispagestyle{empty}

\begin{center}

\vspace{2.5cm}
{\bf HIGH ENERGY SCATTERING IN THE QUASI-POTENTIAL APPROACH }\\

\vspace{0.5cm}

 Nguyen Suan Han $^{a,b,}$\footnote{Email: lienbat76@gmail.com}, Le Hai Yen $^{a}$,  Nguyen Nhu Xuan $^{c}$\\
 \vspace{0.5cm}
{$^a$\it Department of Theoretical Physics,Vietnam National University, Hanoi, Vietnam.} \\
{$^b$\it The Abdus Salam International Centre for Theoretical Physics, Trieste, Italy.}\\
{$^c$\it Department of Theoretical Physics, Le Qui Don Technical University, Hanoi, Vietnam.}\\%

\end{center}

\vspace{0.5cm}
\centerline{\bf Abstract}
\baselineskip=18pt
\bigskip
Asymptotic behavior of the scattering amplitude for  two
scalar particles by scalar, vector and tensor exchanges  at high
energy and fixed momentum transfers is reconsidered in quantum field
theory. In the framework of the quasi-potential approach and the
modified perturbation theory a systematic scheme of finding the
leading eikonal scattering amplitudes and its corrections are
developed and constructed.The connection between the solutions
obtained by quasi-potential and functional approaches is also
discussed.The first correction to leading eikonal amplitude is
found.\\



\newpage

\section{Introduction}

The eikonal scattering amplitude  for the high energy of the two
particles in the limit of high energies and fixed momentum transfers
is found by many authors in quantum field theory $[1-9]$, including
the quantum gravity $[9-20]$. Comparison of the results obtained by
means of the different approaches for this problem has shown that
they all coincide in the leading order approximation, while the
corrections (non-leading terms) provided by them are rather
different $ [15,17,20,21,22,23]$. Determination of these corrections
to gravitational scattering is now open problem $[10-14]$. These
corrections play crucial role in  such problems like  strong
gravitational forces near black hole, string modification
of theory of gravity and other effects of quantum gravity $[9-20]$.\\

The purpose of the present paper is to develop a systematic scheme
based on modified perturbation theory to find the correction terms
to the leading eikonal amplitude for high-energy scattering by means
of solving the Logunov-Tavkhelidze quasi-potential equation
$[24-27]$. In spite of the lack of a clear relativistic covariance,
the quasi-potential method keeps all information about properties of
scattering amplitude which could be received starting from general
principle of quantum field theory $[25]$. Therefore, at high
energies one can investigate the analytical properties of the
scattering amplitude, its asymptotic behavior and some regularities
of a potential scattering etc. Exactly, as it has been done in the
usual S-matrix theory $[24]$. The choice of this approach is
dictated also by the following reasons: 1. in the framework of the
quasi-potential approach the eikonal amplitude has a rigorous
justification in quantum field theory [4]; 2. in the case of smooth
potentials, it was shown that a relativistic quasi-potential and the
Schrodinger
equations lead to qualitatively identical results $[28,29]$.\\

The outline of the paper is as follows. In the second section the
Logunov-Tavkhelidze quasi-potential equation is written in an
operator form. In the third section  the solution of this equation
is presented in an exponent form which is favorable to modify the
perturbation theory. The asymptotic behavior scattering amplitude at
high energies and fixed momentum transfers is considered in the
fourth section. The lowest-order approximation of the modified
theory is the leading eikonal scattering amplitude. Corrections to
leading eikonal amplitude are also calculated. In the fifth section
the solution of quasi-potential equation is presented in the form of
a functional path integral. The connection between the solutions
obtained by quasi-potential and functional integration  is also
discussed. It is shown that the approximations used are similar and
the expressions for corrections to the leading eikonal amplitude are
found identical. Finally, we draw our conclusions.
\section{Two particle quasi-potential equation}
For simplicity, we shall first consider the elastic scattering of
two scalar nucleons interacting with  a scalar meson fields the
model is described by the interaction Lagrangian
$L_{int}=g\varphi^{2}(x)\phi(x)$. The results will be generalized to
the case of scalar nucleons interacting with a neutral vector and
graviton fields later. Following Ref.$[23]$ for two scalar particle
amplitude the quasi-potential equation with local quasi-potential
has the form:
$$T({\bf{p}},{\bf{p}'};s)=gV({\bf{p}}-{\bf{p}'};s)+g\int  d{\bf{q}} V({\bf{p}}-{\bf{q}};s) K({\bf{q}}^{2},s)
T({\bf{q}},{\bf{p'}};s), \eqno(2.1)$$ \\
where
$K({\bf{q}}^{2},s)=\frac{1}{\sqrt{q^2+m^2}}\frac{1}{q^2+m^2-\frac{s}{4}-i\varepsilon}
$, $s=4({\bf{p}}^{2}+m^{2})=4({\bf{p'}}+m^{2})$ is the energy and
${\bf{p}}, {\bf{p'}}$ and are the relative momenta of particles in
the center of mass system in the initial and final states
respectively. Equation $(2.1)$ is one of the possible
generalizations  of the Lippman-Schwinger equation for the case of
relativistic quantum field theory. The quasi-potential $ V $ is a
complex function of the energy and the relative momenta. The
quasi-potential  equation simplifies considerably if $ V $ is a
function of only the difference of the relative momenta and the
total energy, i.e. if the quasi-potential is local \footnote{Since
the total energy $ s $ appears as an external parameter of the
equation, the term "local" here has direct meaning and it can appear
in a three-dimensional $\delta $-function in the quasi-potential in
the coordinate representation }. The existence of a local
quasi-potential has been proved rigorously in the weak coupling case
$[27]$ and a method has been specified for constructing it. The
local potential constructed in this manner gives a solution of Eq.
$(2.1)$, being equal to the physical amplitude on the mass shell $ [24-26]$ .\\
Making the following Fourier transformations
$$
V({\bf{p}}-{\bf{p'}};s)=\frac{1}{(2\pi)^{3}}\int d{\bf{r}}
e^{i({\bf{p}}-{\bf{p}'}){\bf{r}}} V({\bf{r}};s),\eqno (2.2)$$
$$
T({\bf{p}},{\bf{p'}};s)=\int d{\bf{r}} d {\bf{r'}}
e^{i({\bf{p}}{\bf{r}}-{\bf{p'}}{\bf{r'}})} T({\bf{r}},{\bf{r'}};s).
\eqno(2.3)$$
Substituting $(2.2)$ and $(2.3)$ into $(2.1)$ , we obtain
$$
T(\bf r, \bf r' ; s)=\frac{g}{(2\pi)^{3}}V(\bf r; s)\delta^{(3)}(\bf
r- \bf r')+ $$
$$\frac{g}{(2\pi)^{3}}\int\int d\bf q K( \bf q^{2}; s)
V(\bf r; s) e^{-\bf q \bf r} \int d\bf r'' e^{i\bf q \bf r''} T( \bf
r'', \bf r' ; s) \eqno (2.4)
$$
and introducing the representation
$$
T({\bf{r}},{\bf{r}'};s)= \frac{g}{(2\pi)^{3}}
V({\bf{r}};s)F({\bf{r}},{\bf{r'}};s), \eqno(2.5)$$ we obtain
$$
F({\bf{r}},{\bf{r'}};s)=\delta^{(3)}
({\bf{r}}-{\bf{r'}})+\frac{g}{(2\pi)^{3}}\int d{\bf{q}}
K({\bf{q}}^{2};s) e^{-i{\bf{q}}{\bf{r}}}\times \int d
{\bf{r''}}e^{i{\bf{q}}{\bf{r''}}}
V({\bf{r''}};s)F({\bf{r''}},{\bf{r'}};s). \eqno(2.6)
$$
Defining the pseudo-differential  operator
$$
\widehat{L_{r}}=K( - {\bf{{\nabla_{r}}}}^{2};s), \eqno(2.7)
$$
then
$$
K({\bf{r}};s)=\int d{\bf{q}}
K({\bf{q}}^{2};s)e^{-i{\bf{q}}{\bf{r}}}=K( -\nabla_{r}; s) \int
d{\bf{q}} e^{-i{\bf{q}}{\bf{r}}} = \widehat{L_{r}}
(2\pi)^{3}\delta^{(3)}({\bf{r}}). \eqno(2.8)
$$
With allowance for $(2.7)$ and $(2.8)$, Eq. $(2.6)$ can be rewritten
in the symbolic form:
$$
F({\bf{r}},{\bf{r'}};s)=\delta^{(3)}({\bf{r}}-{\bf{r'}})
+g\widehat{L_{r}}\bigl[V({\bf{r}},s)F({\bf{r}},{\bf{r'}},s)\bigl].
\eqno(2.9)
$$
Eq. $(2.8)$ is the Logunov-Tavkhelize quasi-potential equation in
the operator form.
\section{ Modified perturbation theory}

In the framework of the quasi-potential equation the potential is
defined as an infinite power series in the coupling constant which
corresponds to the perturbation expansion of the scattering
amplitude on the mass shell. The approximate equation has been
obtained only in the lowest order of the quasi-potential. Using this
approximation the relativistic eikonal expression of elastic
scattering amplitude was first found in quantum field theory for
large energies and fixed  momentum transfers $[22]$. In this paper
we follow a somewhat different approach based  on the idea of the
modified perturbation theory proposed by Fradkin$[30]$.
\footnote{The interpretation of the perturbation theory from the
view-point of the diagrammatic technique is as follows. The typical
Feynman denominator of the standard perturbation theory is of the
form $(A)$: $(p+\sum q_i )^{2}+m^{2}-i\varepsilon=
p^{2}+m^{2}+2p\sum q_i +(\sum q_i)^{2}$, where $ p $ is the external
momentum of the scalar (spinor) particle, and the $ q_i$ are virtual
momenta of radiation quanta. The lowest order approximation $ (A) $
of modified theory is equivalent to summing all Feynman diagrams
with the replacement: $ (\sum q_i)^{2}=\sum (q_i)^{2} $ in each
denominator $(A)$. The modified perturbation theory thus corresponds
to a small correlation of the radiation quanta: $
\bf{q_{i}}\bf{q_{j}}=0$ and is often called the
$\bf{q_{i}}\bf{q_{j}}$-approximation. In the framework of functional
integration this approximation is called the straight-line path
approximation i.e high-energy particles move along Feynman paths,
which are practically rectilinear [18,19].} The solution of equation
$(2.8)$ can be found in the form

$$
F({\bf{r}},{\bf{r'}};s)=\frac{1}{(2\pi)^3}\int d {\bf{k}}
\exp{\biggl[W({\bf{r}};{\bf{k}};s)\biggl]}
e^{-i{\bf{k}}({\bf{r}}-{\bf{r'}})}. \eqno(3.1)
$$
Substituting $(3.1)$ into $(2.9)$ we have
$$
\exp {W (\bf{r}, \bf{k};s)}= 1 +
g\Bigr\{\widehat{L_{r}}\left[V(\bf{r}; s)\exp{W(\bf{r},
\bf{k};s)}\right]
$$
$$+V(\bf{r};s)\exp{W(\bf{r},\bf{k};s)}K(\bf{k}^{2};s)\Bigr\}. \eqno
(3.2)$$ Reducing this equation for the function $
W({\bf{r}};{\bf{k}};s)$, we get
$$
\exp{W({\bf{r}};{\bf{k}};s)}=1+g \widehat{L_{r}}\left\{V({\bf{r}},s)
\exp \left[W(\bf{r},\bf{k};s)-i\bf{k}\bf{r}\right]\right\}
e^{i{\bf{k}}{\bf{r}}}. \eqno (3.3)
$$
The function $W({\bf{r}};{\bf{k}};s)$ in exponent $ (3.1) $ can now
be written as an expansion in series in the coupling constant g:
$$
W({\bf{r}};{\bf{k}};s)= \sum_{n=1}^{\infty} g^{n}
W_{n}({\bf{r}};{\bf{k}};s). \eqno(3.4)
$$
Substituting $(3.4)$ into $(3.3)$ and using Taylor expansion, the
lhs.$(3.3)$ is rewritten as follow
$$1+\sum_{n=1}^\infty g^nW_n+\frac{1}{2!}\left(\sum_{n=1}^\infty
g^nW_n\right)^2+\frac{1}{3!}\left(\sum_{n=1}^\infty
g^nW_n\right)^3+\ldots,\eqno(3.5)$$ and the rhs.$(3.3)$ has form
$$
1+g\Biggr\{\hat{L}_r\Bigr[V(\mathbf{r};s)\bigr(1+\sum_{n=1}^\infty
g^nW_n +\frac{1}{2!}\left(\sum_{n=1}^\infty
g^nW_n\right)^2+\frac{1}{3!}\left(\sum_{n=1}^\infty
g^nW_n\right)^3+\ldots\bigr)\Bigr]+$$
$$
+V(\mathbf{r};s)\Bigr[1+\sum_{n=1}^\infty g^nW_n
+\frac{1}{2!}\left(\sum_{n=1}^\infty
g^nW_n\right)^2+\frac{1}{3!}\left(\sum_{n=1}^\infty
g^nW_n\right)^3+\ldots\Bigr]K(\mathbf{k};s)\Biggr\}.\eqno(3.6)
$$
From $(3.5)$ and $(3.6)$, to compare with two sides  of Eq.$ (3.3)$
following $ g $ coupling, we derive the following expressions for
the functions $W_{n}({\bf{r}};{\bf{k}};s)$
$$
W_{1}({\bf{r}};{\bf{k}};s)= \int d{\bf{q}}V({\bf{q}};s)
 K[({\bf{k}}+{\bf{q}})^{2};s] e^{-i{\bf{q}}{\bf{r}}}; \eqno(3.7)
$$
$$
W_{2}({\bf{r}};{\bf{k}};s)= -\frac{W_{1}^{2}({\bf{r}};{\bf{k}};s)}
{2!}+ \frac{1}{2} \int d{\bf{q}}_{1} d{\bf{q}}_{2}
V({\bf{q}}_{1};s)V({\bf{q}}_{2};s) K[({\bf{k}}+ {\bf{q}}_{1}+
{\bf{q}}_{2})^2;s]\times
$$
$$\times \left[K (\bf{k}+ \bf{q}_1; s)
+K(\bf{k}+\bf{q}_2;s)\right] e^{
-i{\bf{q}}_{1}{\bf{r}}-i{\bf{q}}_{2}\bf{r}}; \eqno(3.8)
$$
$$
W_{3}({\bf{r}};{\bf{k}};s)=
-\frac{W_{1}^{2}({\bf{r}};{\bf{k}};s)}{3!}+ \int
d{\bf{q}}_{1}d{\bf{q}}_{2}d{\bf{q}}_{3}
V({\bf{q}}_{1};s)V({\bf{q}}_{2};s)V({\bf{q}}_{3};s)
K[({\bf{k}}+{\bf{q}}_{1})^{2};s]$$
$$\times K[({\bf{k}}+
{\bf{q}}_{1}+{\bf{q}}_{2})^{2};s] K[({\bf{k}}+
{\bf{q}}_{1}+{\bf{q}}_{2}+{\bf{q}}_{3})^{2};s ]
e^{-i(\bf{q}_1+\bf{q}_2+\bf{q}_3)\bf{r}}. \eqno(3.9)
$$

Oversleeves  by $ W_{1} $ only we obtain from Eqs $( 3.1)$, $(3.4)$
and $(2.3)$ the approximate expression for the scattering amplitude
[22]
$$
T_{1}({\bf{p}},{\bf{p'}};s)=\frac{g}{(2\pi)^{3}} \int d{\bf{r}}
e^{i({\bf{p}}-{\bf{p'}}){\bf{r}}} V({\bf{r}},s)
e^{gW_{1}({\bf{r}},{\bf{p}},s)}. \eqno(3.10)
$$

To establish the meaning of this approximation, we expand $T_{1}$ in
a series in $g$:

$$
T_{1}^{(n+1)}({\bf{p}},{\bf{p'}};s)=\frac{g^{n+1}}{n!}\int
d{\bf{q}}_{1}...d{\bf{q}}_{n} V({\bf{q}}_{1};s)....V({\bf{q}}_{n};s)
$$
$$\times V({\bf{p}}-{\bf{p'}} -\sum_{i=1}^{n}
q_{i};s) \prod_{i=0}^{n} K[({\bf{q}}_{i}+{\bf{p'}})^{2};s].
\eqno(3.11)
$$

Let us compare Eq. $(3.10)$ with the $ (n+1)-th $ iteration term of
exact Eq. $(2.1)$

$$
T^{(n+1)}(\bf{p},\bf{p'};s)=\int d\bf{q}_1\ldots d\bf{q}_n
V(\bf{q}_1;s)\ldots V(\bf{q}_n;s)\times V(\bf{p}-\bf{p'}
-\sum_{i=1}^n q_i;s)$$
$$ \sum_{p}K[({\bf{q}}_{1}+{\bf{p'}})^{2};s]
K[({\bf{q}}_{1}+ {\bf{q}}_{2}+{\bf{p'}})^{2};s]\ldots K[(
\sum_{i=1}{\bf{q}}_{i}+{\bf{p'}})^{2};s], \eqno(3.12)
$$
where $\sum_{p}$ is the sum over the permutations of the momenta
${\bf{p}}_{1}$ ,${\bf{p}}_{2}...$ ${\bf{p}}_{n}$. It is readily seen
from $(3.11)$ and $(3.12)$ that our approximation in the case of the
Lippmann-Schwinger equation is identical with the
${{\bf{q}}_{i}}{{\bf{q}}_{j}}$ approximation.\\
\section{Asymptotic behavior of the scattering amplitude at high energies}
In this section  the solution of the Logunov-Tavkhelidze
quasi-potential equation obtained in the previous section for the
scattering amplitude can be used to find the asymptotic behavior as
$ s\rightarrow \infty $ for fixed $ t $. In the asymptotic
expressions we shall retain both the principal term and the
following term, using the formula

$$
e^{W({\bf{r}},{\bf{p'}}; s)}= e^{W_{1}({\bf{r}},{\bf{p'}}; s)}\biggl
[1+g^{2}W_{2}({\bf{r}},{\bf{p'}};s)+...\biggl], \eqno(4.1)
$$
where $ W_{1}$ and $ W_{2}$ are given by $(3.7)$ and $(3.8)$.\\
We take the $ z $ axis along the vector  $ ({\bf{p}}+{\bf{p'}}) $
then

$$
{\bf{p}}-{\bf{p'}}={\bf{{\Delta}_{\perp}}};\quad
{\bf{\Delta}_{\perp}} {\bf {n}_{z}}=0;\quad t=-
{\bf{\Delta}_{\perp}^{2}}. \eqno(4.2)
$$

Noting

$$
K({\bf{p}}+{\bf{p'}};s)=\frac{1}
{\sqrt{({\bf{p}}+{\bf{p'}})^{2}+m^{2}}}\frac{1}
{({\bf{p}}+{\bf{p'}})^{2}-\frac{s}{4}+m^{2}-i\varepsilon}\Biggr
|_{\begin{array}{l}
     s\rightarrow\infty \\
     t- fixed
   \end{array}
}$$
$$
=\frac{2}{s(q_{z}^{2}-i\varepsilon)} \left[1-\frac{ 3q_{z}^{2}+
{\bf{q}_{\perp}}^{2} + {\bf{q}_{\perp}}{\bf{\triangle}_{\perp}}}
{\sqrt{s}(q_{z}-i\epsilon)}\right] +
 O\left(\frac{1}{s^{2}}\right), \eqno(4.3)
$$

and using  Eqs $(3.4)$, $(3.7)$ and $(3.8)$ we obtain

$$
W_{1}=\biggl (\frac{W_{10}}{s}\biggl )+ \biggl
(\frac{W_{11}}{s\sqrt{s}}\biggl ) + O \biggl(\frac{1}{s^{2}}\biggl
); \eqno(4.4)
$$

$$
W_{2}=\biggl (\frac{W_{20}}{s^{2}\sqrt{s}}\biggl ) +O \biggl
(\frac{1}{s^{3}}\biggl ), \eqno(4.5)
$$
where

$$
W_{10}=2\int d{\bf{q}} V({\bf{q}};s) \frac{e^{i{\bf{q}}{\bf{r}}}}
{(q_{z}^{2}-i\varepsilon)^{2}}= 2i\int_{-\infty}^{z} dz'
V\left(\sqrt{{\bf{q}_{\perp}}^{2}+z'^{2}};s\right); \eqno (4.6) $$

$$
W_{11}=-2 \int d\bf{q}V(\bf{q};s) e^{-i\bf{qr}}\frac{3q_z^2+
\bf{q}_\perp^2 + \bf{q}_\perp\bf{\triangle}_\perp}
{(q_z-i\epsilon)^2}$$
$$ =-6 V\left(\sqrt{\bf{q}_\perp^2+z'^2};s\right) + 2(-\bf{\nabla}_\perp^2-
i\bf{q}_\perp\bf{\nabla}_\perp)\times\int_{-\infty}^z dz'
V\left(\sqrt{\bf{q}_\perp^2+z'^2};s\right); \eqno(4.7)$$

$$
W_{20}=-4 \int d\bf{q}_1d\bf{q}_2 e^{-i(\bf{q}_1+\bf{q}_2)\bf{r}}
V(\bf{q}_1;s)V(\bf{q}_2;s)$$
$$
\times \frac{3q_{1z}q_{2z}+\bf{q}_{1\perp}\bf{q}_{2\perp}}
{(q_{1z}-i\varepsilon)(q_{2z}-i\varepsilon)(q_{1z}+q_{2z}-i\varepsilon)}$$
$$= -4i\Biggr\{ 3 \int_{-\infty}^z dz'
V^2\left(\sqrt{\bf{q}_\perp^2+z'^2};s\right)+\left[{\bf{{\nabla}_{\perp}}}\int_{-\infty}^{z'}
dz''
V^2\left(\sqrt{{\bf{q}_{\perp}}^{2}+z''^{2}};s\right)\right]^{2}\Biggr\}.
\eqno(4.8)
$$

In the limit $s\rightarrow \infty $ and $ (t/s) \rightarrow 0 $ $
W_{10}$ makes the main contribution, and the remaining terms are
corrections. Therefore, the function $ \exp {W}$ can be represented
by means of the expansion $(4.1)$ where $W_{10}$, $ W_{11}$ and $
W_{20}$ are determined by Eqs. $(4.6)-(4.8)$ respectively. The
asymptotic behavior scattering amplitude can be written in the
following form

$$
T(\bf{p},\bf{p'}; s)=\frac{g}{(2\pi)^{3}}\int d^{2}\bf{r}_{\perp}dz
e^{i{\bf{\Delta_{\perp}}}{\bf{r_{\perp}}}}V\left(\sqrt{\bf{r}^{2}+z^{2}};s\right)$$
$$\times \exp\biggl(g\frac{W_{10}}{s}\biggl)
\biggl(1+g\frac{W_{11}}{s\sqrt{s}}+g^{2}\frac{W_{20}}{s^{2}\sqrt{s}}+\ldots\biggl).
\eqno(4.9)
$$

Substituting $(4.6)-(4.8)$ into $(4.9)$ and making calculations, at
high energy $s\rightarrow\infty$ and fixed momentum transfers
$(t/s)\rightarrow 0 $, we finally obtain[22]
$$
T(s,t)=\frac{g}{2i(2\pi)^3} \int d^2 \bf{r}_\bot
e^{i\bf{\Delta}_\bot\bf{r}_\bot}\times\Biggl\{e^{\Bigl[\frac{2ig}{s}\int_{-\infty}^\infty
dz V\left(\sqrt{\bf{r}^2_\bot+z^2};s\right)\Bigl]}-1\Biggl\}
$$
$$
-\frac{6g^{2}}{(2\pi)^{3}s\sqrt{s}} \int
d^{2}\bf{r}_{\perp}e^{i\bf{\Delta}_\bot\bf{r}_\bot}\times\exp\biggl[\frac{2ig}{s}
\int_{-\infty}^{\infty} dz'
V\left(\sqrt{\bf{r}_{\perp}^{2}+z^{2}};s\right)\biggl]$$
$$\times\int_{-\infty}^{\infty} dz
V\left(\sqrt{\bf{r}_{\perp}^{2}+z^{2}};s\right)
-\frac{ig}{(2\pi)^3\sqrt{s}}\int d^2\bf{r}_\bot
e^{i\bf{\Delta}_\bot\bf{r}_\bot}\times$$
$$\int_{-\infty}^{\infty}dz \Biggl\{\exp \biggl[\frac{2ig}{s}
\int_{z}^\infty
dz'V\left(\sqrt{\bf{r}_\bot^2+z'^2};s\right)\biggl]-\exp{\biggl[\frac{2ig}{s} \int_{-\infty}^{\infty} dz'
V\left(\sqrt{\bf{r}_{\perp}^{2}+z'^{2}};s\right)\biggl]}\Biggl\}$$
$$\times\Biggl\{\int_z^\infty dz'\bf{\nabla}_\bot^2V\left(\sqrt{\bf{r}_\bot^2+z'^2};s\right)- \frac{2ig}{s} \biggl[\int_z^\infty dz'\bf{\nabla}_\bot
V\left(\sqrt{\bf{r}_\bot^2+z^2};s\right)\biggl]^2\Biggl\}$$
$$-\frac{2ig}{(2\pi)^3s}\bf{\Delta}_\perp^2
\int
d^{2}\bf{r}_{\perp}V\left(\sqrt{\bf{r}_{\perp}^{2}+z'^{2}};s\right)]
e^{i{\bf{\Delta_{\perp}}}{\bf{r_{\perp}}}}\times$$
$$
\int_{-\infty}^{\infty} zdz V(\sqrt{\bf{r}_{\perp}^{2}+z^{2}};s)
\exp {\biggl[ \frac{2ig}{s} \int_{z}^{\infty}
dz'V(\sqrt{\bf{r}_{\perp}^{2}+z'^{2}};s)\biggl ]}. \eqno(4.10)
$$

In this expression $(4.10)$ the first term describes the leading
eikonal behavior of the scattering amplitude, while the remaining
terms determine the corrections of relative magnitude $1/\sqrt{s}$.
The similar result Eq.$(4.10)$ is also found  by means of the
functional integration $[20]$.\\

As is well known from the investigation of the scattering amplitude
in the Feynman diagrammatic technique, the high-energy asymptotic
behavior can contain only logarithms and integral powers of $ s $. A
similar effect is observed here, since integration of the expression
$(4.10)$  leads to the vanishing of the coefficients for
half-integral powers of $s$. Nevertheless, allowance for the terms
that contain the half-integral powers of $s$ is needed for the
calculations of the next corrections in the scattering amplitude,
and leads to the appearance of the so-called retardation effects,
which are absent in the  principal asymptotic term.\\

    In the limit of high energies $s\rightarrow \infty $ and
for fixed momentum transfers $t$ the expression for the scattering
amplitude within the framework of the functional - integration
method takes the Glauber form with eikonal function corresponding to
a Yukawa interaction potential between "nucleons". Therefore,  the
local quasi-potential for the interaction between the "nucleons"
from perturbation theory in that region can be chosen by following
forms. For the scalar meson exchange the quasi-potential decreases
with energy

$$
V(r;s)=-\frac{g^2}{8\pi s}\frac{e^{-\mu r}}{r}. \eqno(4.11)
$$

The first term  in the expression $(4.10)$ describes the leading
eikonal behavior of the scattering amplitude. Using integrals
calculated in the Appendix, we find

$$
T^{(0)}_{Scalar}(s,t)=-\frac{g}{2i(2\pi)^3}\int d^2\bf{r}_\bot
e^{i\Delta_\bot \bf{r}_\bot}\times$$
$$\left\{\exp\left[\frac{2ig}{s}\int_{-\infty}^{+\infty} dz
V\left(\sqrt{\bf{r}_\bot^2+z^2};s\right)\right]-1\right\} $$
$$
=\frac{g^4}{4(2\pi)^4s^2}\left[\frac{1}{\mu^2-t}-\frac{g^3}{8(2\pi)^2
s^2}F_{1}(t)+\frac{g^6}{48(2\pi)^5s^4}F_2(t)\right]. \eqno(4.12)$$

The next term in $(4.10)$ describes first correction to the leading
eikonal amplitude

$$
T^{(1)}_{Scalar}(s,t)=-\frac{6g^2}{(2\pi)^3s\sqrt{s}}\int d^2r_\bot
e^{i\Delta_\bot r_\bot}\times$$
$$\exp\left[\frac{2ig}{s}\int_{-\infty}^{+\infty} dz
V\left(\sqrt{\bf{r}_\bot^2+z^2};s\right)\right]
\times\int_{-\infty}^{+\infty} dz
V\left(\sqrt{\mathbf{r}_\bot^2+z^2};s\right)$$
$$=\frac{3g^4}{4(2\pi)^6s^2\sqrt{s}}\left[\frac{2}{\mu^2-t}
-\frac{g^3}{2(2\pi)^2s^2}F_{1}(t)+\frac{g^6}{8(2\pi)^5s^4}F_{2}(t)\right]\eqno(4.13)
$$
where
$$
F_1(t)= \frac{1}{t\sqrt{1-\frac{4\mu^2}{t}}}ln\left|
\frac{1-\sqrt{1-4\mu^2/t}}{1+\sqrt{1-4\mu^2/t}}\right|\eqno(4.14)
$$
and

$$
F_2(t)=\int_0^1
dy\frac{1}{(ty+\mu^2)(y-1)}ln\left|\frac{\mu^2}{y(ty+\mu^2-t)}\right|\eqno(4.15)
$$

A similar calculations can be applied  for other exchanges with
different spins. In the case of the vector model $ L_{int}=
-g\varphi^{\star}i\partial_\sigma\varphi
A^{\sigma}+g^{2}A_{\sigma}A^{\sigma}\varphi\varphi^{\star}$ the
quasi-potential is independent of energy
$$V(r;s)=-\frac{g^2}{4\pi}\frac{e^{-\mu r}}{r},$$we find

$$
T^{(0)}_{Vector}(s,t)=\frac{g^4}{2(2\pi)^4s}\times\left[\frac{1}{\mu^2-t}-\frac{g^3}{4(2\pi)^2
s}F_{1}(t)+\frac{g^6}{12(2\pi)^5s^2}F_2(t)\right]\eqno(4.16)
$$

$$
T^{(1)}_{Vector}(s,t)=\frac{3g^4}{2(2\pi)^6s\sqrt{s}}\times\left[\frac{2}{\mu^2-t}
      -\frac{g^3}{(2\pi)^2s}F_{1}(t)+\frac{g^6}{2(2\pi)^5s^2}F_{2}(t)\right]\eqno(4.17)
$$

In the case of tensor model \footnote{The model of interaction of a
scalar "nucleons" field $\varphi(x) $ with a gravitational field $
g_{\mu\nu}(x)$ in the linear approximation to $
h^{\mu\nu}(x)$;$[18]$ $ L(x)=L_{0,\varphi}(x)+L_{0,grav.}(x)
+L_{int}(x) $,where

$$ L_0(x)=\frac{1}{2} \left[\partial^{\mu} \varphi (x)\partial_{\mu} \varphi(x)
-m^{2} {\varphi}^{2}(x)\right] ,$$

$$ L_{int}(x)=-\frac{\kappa}{2} h^{\mu\nu}(x)T_{\mu\nu}(x) ,$$

$$ T_{\mu\nu}(x)=\partial_{\mu}\varphi(x) \partial_{\nu}\varphi(x)-
\frac{1}{2}\eta_{\mu\nu} \left[\partial^{\sigma} \varphi
(x)\partial_{\sigma} \varphi(x) -m^{2} {\varphi}^{2}(x)\right ] ,$$

$ T_{\mu\nu}(x)$-the energy momentum tensor of the scalar field. The
coupling constant $\kappa$ is related to Newton's constant of
gravitation $ G $ by $ \kappa^2=16 \pi G $ }, the quasi- potential
increases with energy $V(r; s)= (\kappa^2 s/2\pi)(e^{-\mu r}/r)$, we
have

$$
T^{(0)}_{Tensor}(s,t)=-\frac{\kappa^4}{(2\pi)^4}\times\left[\frac{1}{\mu^2-t}
  +\frac{\kappa^3}{2(2\pi)^2}F_{1}(t)+\frac{\kappa^6}{3(2\pi)^5}F_2(t)\right]\eqno(4.18)
$$
$$
T^{(1)}_{Tensor}(s,t)=-\frac{3\kappa^4}{(2\pi)^6\sqrt{s}}\times\left[\frac{2}{\mu^2-t}
      +\frac{2\kappa^3}{(2\pi)^2}F_{1}(t)+\frac{2\kappa^6}{(2\pi)^5}F_{2}(t)\right]\eqno(4.19)
$$

 To conclude this section it is important to note that
in the framework of standard field theory for the high-energy
scattering, different methods have been developed to investigate the
asymptotic behavior of individual Feynman diagrams and their
subsequent summation. In different theories including quantum
gravity the calculations of Feynman diagrams in the eikonal
approximation is proceed  in a similar way as analogous the
calculations in QED. Reliability of the eikonal approximation
depends on spin of the exchanges field $[5,6]$. The eikonal captures
the leading behavior of each order in perturbation theory, but the
sum of leading terms is subdominant to the terms neglected by this
approximation. The reliability of the eikonal amplitude for gravity
is uncertain [14]. Instead of the diagram technique perturbation
theory, our approach is based on the exact expression of the
scattering amplitude and modified perturbation theory which in
lowest order contains the leading eikonal amplitude and the next
orders are its corrections.

\section{Relationship between the operator and Feynman path methods }

What actual physical picture may correspond to our result given by
Eq. $(4.10)$ ? To answer  this question we establish the
relationship between the operator and Feynman path methods in Ref.
[31], which treats the quasi-potential equation in the language of
functional integrals. The solution of this equation can be written
in the symbolic form:

$$
\exp (W)=\frac{1}{1-gK[(-i {\bf{{\nabla}}}-{\bf{k}})^{2}] V({\bf
{r}})}\times \bf{1}$$
$$ =-i\int_{0}^{\infty}d\tau \exp [i\tau(1+i\varepsilon)]\times\exp \left\{-i\tau g K[(-i {\bf{{\nabla}}}-{\bf{k}})^{2}]
V({\bf{r}})\right\}\times \bf{1}. \eqno(5.1) $$

In accordance with the Feynman parametrization $[31]$, we introduce
an ordering index $ \eta $ and write Eq. $(5.1)$ in the form

$$
\exp (W)= -i\int_0^\infty d\tau e^{i\tau
(1+i\varepsilon)}\times\exp\left\{ -ig \int_0^\infty d\eta
K[(-i{\bf{{\nabla}_{\eta+\varepsilon}}}-{\bf{k}})^{2}]
U({\bf{r}}_{\eta})\right\} \times \bf{1}. \eqno(5.2)
$$

Using Feynman transformation

$$
F[P(\eta)]= \int D{\bf{p}}\int_{x(0)=0}
\frac{D{\bf{x}}}{(2\pi)^{3}}\times\exp\left\{i \int_{0}^{\tau} d\eta
\dot{{\bf{r}}}(\eta)[ {\bf{p}}(\eta)-P(\eta)]\right\}
F[{\bf{p}}(\eta)], \eqno(5.3)
$$

we write the solution of  Eq. $(2.8)$ in the form of the functional
integral

$$
\exp (W)=-i\int_{0}^{\infty} d\tau e^{i\tau (1+i\varepsilon)}\int
D{\bf{p}}\int_{x(0)=0} \frac{D{\bf{x}}}{(2\pi)^{3}}$$
$$
\times\exp\left\{i \int_{0}^{\tau} d\eta \dot{{\bf{x}}}(\eta)[
{\bf{p}}(\eta)-P(\eta)]\right\} G({\bf{x}},{\bf{p}};\tau)\times
\bf{1}.\eqno(5.4)
$$

In Eq.$(5.4)$ we enter the function $G $

$$
G(\bf{x},\bf{p};\tau)=\exp \left\{-i \int_0^\tau d\eta
\dot{\bf{x}}(\eta) \nabla_{\eta+\varepsilon}\right\}\times
\exp\left\{ -ig \int_{0}^{\tau} d\eta
K[({\bf{p}}(\eta)-{\bf{k}})^{2}]V({\bf{r}}_{\eta})\right\}, \eqno
(5.5)
$$

which satisfies the equation

$$
\frac{dG}{d\tau}= \left\{ -igK
[({\bf{p}}(\tau)-{\bf{k}})^{2}]V({\bf{r}}-
\dot{{\bf{x}}}(\tau-\varepsilon)){\bf{{\nabla}}}\right\}G;\quad
G(\tau=0)=1. \eqno(5.6)
$$

Finding from  Eq. $(5.6)$  the operator function $ G $ and
substituting it into Eq. $(5.6)$ for $ W $ we obtained the following
final expression:

$$
\exp (W)=-i\int_{0}^{\infty} d\tau e^{i\tau (1+i\varepsilon)}\int D
{\bf{p}} \frac{1}{(2\pi)^{3}}$$

$$\times\int_{x(0)=0}
\frac{D{\bf{x}}}{(2\pi)^{3}} \exp\left\{i \int_{0}^{\tau} d\eta
\dot{{\bf{x}}}(\eta)p(\eta)\right\} \exp (g \prod), \eqno (5.7)
$$

where

$$
\prod =-i\int_{0}^{\infty} d\tau K
[({\bf{p}}(\eta)-{\bf{k}})^{2}]\times
V\left[{\bf{r}}-\int_{0}^{\tau} d\xi \dot{{\bf{x}}(\xi)}\vartheta
(\xi-\eta+\varepsilon)\right]; \eqno (5.8)
$$

$$
{\prod}^{2}=-\int_{0}^{\tau_{1}}\int_{0}^{\tau_{2}}d\tau_{1}d\tau_{2}
K [({\bf{p}}(\eta_{1})-{\bf{k}})^{2}]K
[({\bf{p}}(\eta_{2})-{\bf(k)})^{2}]$$
$$
\times V\left[{\bf{r}_{1}}-\int_{0}^{\tau_{1}} d\xi
\dot{{\bf{x}}(\xi)}\vartheta (\xi-\eta+\varepsilon)\right]\times
V\left[{\bf{r}_{2}}-\int_{0}^{\tau_{2}} d\xi
\dot{{\bf{x}}(\xi)}\vartheta
(\xi-\eta+\varepsilon)\right].\eqno(5.9)
$$

Writing out the expansion $[2,3]$

$$
\exp(W)=\overline{\exp(g\prod)}= \exp({g\overline{\prod}})
\sum_{n=0}^{\infty}\frac{g^{n}}{n!}\overline{\Bigl(\prod-\overline{\prod}\Bigl)^{n}},
$$
in which the sign of averaging denoted integration with respect to $
\tau $, $ \bf{x}(\eta)$ and $ \bf{p}(\eta)$ with the corresponding
measure ( see, for example Eq. $(5.7)$ ), and performing the
calculations, we find

$$
W_{1}=\overline{ \prod },
W_{2}=\frac{\overline{{\prod}^{2}}- {\overline{ \prod }}^{2} }
{2!}, W_{3}=\frac{1}{3!}\Bigl[
\overline{{\prod}^{3}}-{\overline{\prod}}^{3}-
3\overline{\prod}(\overline{{\prod}^{2}}-{\overline{\prod}}^{2})\Bigl],
etc. \eqno(5.10)
$$
i.e. the expressions $(5.10)$ and $(4.1)$ are identical.
$$
W_{1}=\overline{ \prod }=-i\int_0^\infty d\tau
K[(\bf{p}(\eta)-\bf{k})^2]\times \exp \left[-\int_{0}^{\tau} d\xi
\dot{{\bf{x}}(\xi)}\vartheta(\xi-\eta+\varepsilon)\nabla_{\eta}\right]V(\vec{r})
$$
$$
=\int d{\bf{q}} e^{-{\bf{q}}{\bf{r}}} K[({\bf{q}}+{\bf {k}})^{2}]
V({\bf{q}};s); \eqno (5.11)$$

$$ \overline{{\prod}^{2}}=K[({\bf{\nabla}_{\bf{r_{1}}}}+
\bf{\nabla}_{\bf{r_{2}}}+\bf{k})^{2}]
K[(\bf{\nabla}_{\bf{r_{1}}}+\bf{k})^{2}]\times
K[(\bf{\nabla}_{\bf{r_{2}}}+\bf{k})^{2}] V(\bf{r_{1}};s)
V(\bf{r_{2}};s)
$$
$$ =\int d {\bf{q_{1}}}\int d {\bf{q_{2}}}
e^{-i({\bf{q_{1}}}+{\bf{q_{2}}}){\bf{r}}}
K[({\bf{q_{1}}}+{\bf{q_{2}}}+{\bf{k}})^2]$$
$$\times\left\{K[({\bf{q_{1}}}+{\bf{k}})^{2}]+
K[({\bf{q_{2}}+{\bf{k}}})^{2}]\right\} V({\bf{r_{1}}};s)
V({\bf{r_{2}}};s); \eqno(5.12)
$$
$$
W_{2}=-\frac{W_{1}^{2}}{2!}+ \frac{1}{2}\int
d{\bf{q_{1}}}d{\bf{q_{2}}} V({\bf{q_{1}}})V({\bf{q_{2}}})\times\left\{K[({\bf{q_{1}}}+{\bf{k}})^{2};s]+
K[({\bf{q_{2}}}+{\bf{k}})^{2};s]\right\}; \eqno(5.13)
$$

$$
W_{3}=-\frac{W_{1}^{3}}{3!} +  \int
d{\bf{q}}_{1}d{\bf{q}}_{2}d{\bf{q}}_{3}
V({\bf{q}}_{1};s)V({\bf{q}}_{2};s)V({\bf{q}}_{3};s)$$
$$\times K[(\bf{k}+\bf{q}_1)^2;s]K[({\bf{k}}+
{\bf{q}}_{1}+{\bf{q}}_{2})^{2};s] K[({\bf{k}}+
{\bf{q}}_{1}+{\bf{q}}_{2}+{\bf{q}}_{3})^{2};s ]$$
$$\times e^{-i(\bf{q}_1+\bf{q}_2+\bf{q}_3)\bf{r}}; etc \eqno(5.14)
$$

Restricting ourselves in the expansion $(5.10)$ to the first term
$(n=0) $, we obtain the approximate expression $(4.12)$ for the
scattering amplitude, which corresponds to the allowance for the
particle Feynman paths. These paths can be considered as a classical
paths and coincide in the case of the scattering of high-energy
particles through small angles to straight-line paths trajectories.

\section{Conclusions}
Asymptotic behavior of scattering amplitude for  two scalar
particles  at high energy and fixed momentum transfers was studied.
In the framework of quasi-potential approach and the modified
perturbation theory the systematic scheme of finding the corrections
to the principal asymptotic leading scattering amplitudes was
constructed and developed. Results obtained by two different
approaches (quasi potential and functional) for this problem, as it
has shown that they are identical. Results obtained by us are
extended to the case of scalar particles of the field $\varphi(x)$
interacting with a  vector and gravitational fields. The first
correction to the leading eikonal scattering amplitude in quantum
field theory was obtained.
\section*{Acknowledgments}
We are grateful to Profs. B.M. Barbashov,  V.N.Pervushin for
valuable discussions and Prof. G. Veneziano for suggesting this
problem and his encouragement. N.S.H. is also indebted to Prof. H.
Fried for reading the manuscript and making useful remarks for
improvements. This work was supported in part by the International
Center for Theoretical Physics, Trieste, the Abdus Salam
International Atomic Energy Agency, the United Nations Educational,
Scientific and Cultural Organization, by a grand TRIGA and by the
Vietnam National University under Contract QG.TD.10.02.
\appendix
\section{The kernel of the quasi-potential equation $[25]$}
 We denote by $G(\bf{p},\bf{p'},\varepsilon_p,\varepsilon_q,E)$ the total Green
 function for two particles, where $\bf{p}$ and $\bf{p'}$ are
 the momenta of the initial and final states in c.m.s and $2E=\sqrt{s}$ is
 the total energy.\\
In these notations the Bethe-Salpeter equation is of the form
$$
G(\bf{p},\bf{p'},\varepsilon_p,\varepsilon_{p'},E)=iF(\bf{p},\varepsilon_p,E)
\delta(\bf{p}-\bf{p'})\delta(\varepsilon_p-\varepsilon_{p'})$$
$$+F(\bf{p},\varepsilon_p, E)\int K(\bf{p},\bf{q},\varepsilon_p,\varepsilon, E)
G(\bf{q},\bf{p'},\varepsilon,\varepsilon_{p'}, E) d\bf{q}
d\varepsilon,\eqno(A.1)$$ where
$$iF(\bf{p},\varepsilon_p,
E)=\frac{2}{\pi}D(E+\varepsilon_p,\bf{p})D(E-\varepsilon_p,\bf{p})$$
$$D(E+\varepsilon_p,\bf{p})=\frac{1}{(E+\varepsilon_p)^2-p^2-m^2+i\epsilon}.\eqno(A.2)$$
Now we introduce formally the scattering amplitude $T$ which on the
mass - shell $\varepsilon_p=\varepsilon_{p'}=0, \quad
p^2=p'^2=E^2-m^2$ gives the physical scattering amplitude:

$$G(\bf{p},\bf{p'},\varepsilon_p,\varepsilon_{p'}, E)-iF(\bf{p},\varepsilon_p, E)
\delta(\bf{p}-\bf{p'})\delta(\varepsilon_p-\varepsilon_{p'})= $$
$$=iF(\bf{p},\varepsilon_p,E)T(\bf{p},\bf{p'},\varepsilon_p,\varepsilon_{p'},E)
F(\bf{p'},\varepsilon_{p'},E).\eqno(A.3)$$

Then inserting $(A.3)$ into $(A.1)$, we get for $T$ the equation

$$
T(\bf{p},\bf{p'},\varepsilon_p,\varepsilon_{p'}, E) =
K(\bf{p},\bf{p'},\varepsilon_p,\varepsilon_{p'}, E)$$
$$ +\int d\bf{q} d\varepsilon K(\bf{p},\bf{q},\varepsilon_p,\varepsilon, E)
F(\bf{q},\varepsilon,E)
T(\bf{q},\bf{p'},\varepsilon,\varepsilon_{p'},E).\eqno(A.4)$$

We wish to obtain an equation of the Lippmann - Schwinger type for a
certain function $T(\bf{p},\bf{p'},E)$ which on the mass-shell
$p^2=p'^2=E^2-m^2$ would give the physical scattering amplitude:

$$
T(\bf{p},\bf{p'},E)=V(\bf{p},\bf{p'},E)+\int
d\bf{q}V(\bf{p},\bf{q},E)F(\bf{q},E)T(\bf{q},\bf{p'},E),\eqno(A.5)
$$
where
$$
F(\bf{q},E)=\int d\varepsilon
F(\bf{q},\varepsilon,E)=-\frac{2i}{\pi}\int d\varepsilon
\frac{1}{(E+\varepsilon)^2-p^2-m^2}\times
\frac{1}{(E-\varepsilon)^2-p^2-m^2}$$
 $$=\frac{1}{\sqrt{q^2+m^2}(q^2+m^2-E^2)}.\eqno(A.6)
$$
On the mass-shell, the total energy $E=\frac{\sqrt{s}}{2}$, we
receive the kenel that is brought out in Eq.(2.1)
$$K(q^2;s)\equiv F\left(q,E=\frac{\sqrt{s}}{2}\right)
=\frac{1}{\sqrt{q^2+m^2}(q^2+m^2-\frac{s}{4})}. \eqno{(A.7)}$$
  This can be achieved by a
conventional choice of the potential $ V(\bf{p},\bf{p'},E)$, which
can obviously be made by different methods.There are two methods
that have been suggested for constructing a complex potential
dependent on energy with the help of which one can obtain from an
equation of the Schr\"{o}dinger
type the exact scattering amplitude on the mass - shell. \\
The first method is based on the two-time Green function $ [23]$
which in the momentum space is defined

$$ G(\bf{p},\bf{p'},E)=\int d\varepsilon_p d\varepsilon_{p'}
G(\bf{p},\bf{p'},\varepsilon_p,\varepsilon_{p'},E).\eqno(A.8)$$

Then using $(A.3)$ and $(A.8)$ we can determine the corresponding
off-shell scattering amplitude

$$T_1(\bf{p},\bf{p'},E)=\frac{1}{F(\bf{p},E)F(\bf{p'},E)}\int F(\bf{p},\varepsilon_p,E)
T(\bf{p},\bf{p'},\varepsilon_p,\varepsilon_{p'},E) F(\bf{p'},
\varepsilon_{p'},E) d\varepsilon_p d\varepsilon_{p'}.\eqno(A.9)$$

From expression $(A.9)$ it is directly seen that $T$ on mass-shell
$p^2=p'^2=E^2-m^2 $ coincides with the scattering amplitude $
T(\bf{p},\bf{p'},0,0,E)\equiv T(\bf{p},\bf{p'},E)$. The potential
$V_1$ for Eq.$(A.5)$ is constructed by iteration of Eqs.$(A.4)$,
$(A.5)$ and $(A.9)$. In particular, in the lowest order, we have

$$
V_1(\bf{p},\bf{p'},E)=\frac{1}{F(\bf{p},E)F(\bf{p'},E)}\int
F(\bf{p},\varepsilon_p,E)K(\bf{p},\bf{p'},\varepsilon_p,\varepsilon_{p'},E)
 F(\bf{p'}, \varepsilon_{p'},E) d\varepsilon_p
 d\varepsilon_{p'}.\eqno(A.10)
 $$

The second method consists in constructing the potential $V_2$ for
Eq.$(A.4)$ by means of the scattering amplitude T on the mass-shell
obtained by perturbation theory, e.g. from Eq.$(A.4)$ and the
iterations of Eq.$(A.5)$ accompanied by the transition to the
mass-shell.\\

We write down Eq.$(A.5)$ in the symbolic form $ T_2=V_2+V_2\times
T_2$ and obtain in the lowest orders of $V_2$ the expressions

$$V_2^{(2)}=[T^{(2)}], V^{(4)}=[T^{(4)}]-[V_2^{(2)}\times T_2^{(2)}],$$
$$V_2^{(6)}=[T^{(6)}]-[V_2^{(2)}\times T_2^{(4)}]-[V_2^{(4)}\times
T_2^{(2)}]\ldots,\eqno(A.11)$$ where the square brackets mean here
the transition to the mass-shell. Hence, it follows that in the
second method we get a local potential dependent only on
$(\bf{p}-\bf{p'})^2$ and $E$ and in r-space on r and $E$.\\

We shall consider, as an example, the application of the above
methods to a model of quantum field theory, in which scalar
particles of mass m interact by exchanging scalar "photons" of small
mass $\mu$. We shall put $\mu=0$ where it is possible. In the ladder
approximation (without considering the crossing symmetry) the kernel
of Eq.$(A.5)$ is of the form

$$K(\bf{p},\bf{p'},\varepsilon_p,\varepsilon_{p'},E)= -\frac{ig^2}{(2\pi)^4}\frac{1}{(\varepsilon_p-\varepsilon_{p'})^2
 -(\bf{p}-\bf{p'})^2-\mu^2+i\epsilon}.\eqno(A.12)$$
\section{Some integrals used in this paper}

We consider the integral

$$
I_1=\int_{-\infty}^\infty dz
V\left(\sqrt{\mathbf{r}_\bot^2+z^2};s\right)= -\frac{g^2}{8\pi
s}\int_{-\infty}^\infty dz\frac{e^{-\mu
\left(\sqrt{\mathbf{r}_\bot^2+z^2}\right)}}
{\sqrt{\mathbf{r}_\bot^2+z^2}} \eqno(B.1)
$$

we have

$$
I_1=-\frac{g^2}{4(2\pi)^4 s}\int d^3p\int_{-\infty}^{+\infty}dz
\frac{e^{i\textbf{pr}}}{\mu^2+p^2}= -\frac{g^2}{4(2\pi)^4 s}\int
d^3p\int_{-\infty}^{+\infty}dz \frac{e^{i(p_\bot r_\bot +
p_{//}z)}}{\mu^2+p^2}$$
$$
=-\frac{g^2}{4(2\pi)^4 s}\int d^3p \frac{e^{i(p_\bot r_\bot
)}}{\mu^2+p^2}\int_{-\infty}^{+\infty}dz e^{ip_{//}z}
$$
$$
=-\frac{g^2}{4(2\pi)^4 s}\int d^2p_\bot dp_{//}\frac{e^{i(p_\bot
r_\bot )}}{\mu^2+p_\bot^2+p_{//}^2}\times(2\pi)\delta(p_{//})
$$
$$
=-\frac{g^2}{4(2\pi)^3 s}\int d^2p_\bot e^{i(p_\bot r_\bot )}\int
dp_{//}\frac{\delta(p_{//})}{\mu^2+p_\bot^2+p_{//}^2}
$$
$$
=-\frac{g^2}{4(2\pi)^3 s}\int d^2 p_\bot \frac{e^{i(p_\bot
r_\bot)}}{\mu^2+p_\bot^2} =-\frac{g^2}{4(2\pi)^2 s}K_0\left(\mu
|r_\bot|\right),\eqno(B.2)
$$
with $K_0\left(\mu |r_\bot|\right)=\frac{1}{2\pi}\int d^2 p_\bot
\frac{e^{i(p_\bot r_\bot)}}{\mu^2+p_\bot^2}$ is the MacDonald
function of zeroth order. \\

 The integral
$$
I_2=\int d^2r_\bot e^{i\Delta_\bot
r_\bot}K_0(\mu|r_\bot|)=(2\pi)\int d |r_\bot||r_\bot| J_{(0)}(\Delta_\bot
|r_\bot|)K_0\left(\mu |r_\bot|\right) =\frac{2\pi}{\mu^2-t}. \eqno(B.3)
$$

The integral
$$
I_3=\int d^2r_\bot e^{i\Delta_\bot r_\bot}K_0^2(\mu|r_\bot|)
=\int d^2 r_\bot e^{i\Delta_\bot r_\bot}\left(\frac{1}{2\pi}\int
d^2q \frac{e^{i\textbf{qr}_\bot}}{q^2+\mu^2}\right)K_0\left(\mu
|r_\bot|\right)
$$
$$
=\frac{1}{2\pi}\int d^2q \frac{1}{q^2+\mu^2}\int d^2 r_\bot
e^{i\left(q+\Delta_\bot\right) r_\bot}K_0\left(\mu|r_\bot|\right)
$$
$$=\frac{1}{2\pi}(2\pi)\int d^2q
\frac{1}{q^2+\mu^2}\frac{1}{\left(q+\Delta_\bot\right)^2
+\mu^2},\eqno(B.4)
$$
here, the result of the integral that obtained from calculating
$I_2$ have been used. \\

Using method of Feynman parameter integral
$\frac{1}{ab}=\int_0^1\frac{dx}{[ax+b(1-x)]^2}$, we have

$$
I_3=\int_0^1 dx\int d^2q \times\frac{1}{\left\{(q^2+\mu^2)x
+\left[\left(q+\Delta_\bot\right)^2+\mu^2\right](1-x)\right\}^2}
$$
$$
=\int_0^1 dx \int d^2q\frac{1}{\left[q^2+2q\Delta_\bot(1-x)
+\Delta_\bot^2(1-x)+\mu^2\right]^2}
$$
$$
=\int_0^1 dx
\frac{i(-\pi)\Gamma(1)}{\left[\Delta_\bot^2(1-x)+\mu^2-\Delta_\bot^2(1-x)^2\right]\Gamma(2)}
$$
$$
=(-i\pi)\int_0^1 dx\frac{1}{\left[\mu^2+\Delta_\bot^2x(1-x)\right]}
=(-i\pi)\int_0^1\frac{dx}{\left[\mu^2-tx(1-x)\right]}=
$$
$$
=(-i\pi)\times\frac{1}{t\sqrt{1-\frac{4\mu^2}{t}}}
ln\frac{1-\sqrt{1-\frac{4\mu^2}{t}}}{1+\sqrt{1-\frac{4\mu^2}{t}}}
\equiv(-i\pi)\times F_{1}(t). \eqno(B.5)
$$

Finally, we calculate the integral

$$
I_4=\int d^2r_\bot e^{i\Delta_\bot r_\bot}K_0^3(\mu|r_\bot|)
$$
$$
=\int d^2r_\bot e^{i\Delta_\bot r_\bot}\left(\frac{1}{2\pi}\int
d^2q_1\frac{e^{iq_1r_\bot}}{q_1^2+\mu^2}\right)\times\left(\frac{1}{2\pi}\int
d^2q_2\frac{e^{iq_2r_\bot}}{q_2^2+\mu^2}\right)K_0(\mu
\left|r_\bot\right|)
$$
$$
=\frac{1}{(2\pi)^2}\int\frac{d^2q_1d^2q_2}{(q_1^2+\mu^2)(q_2^2+\mu^2)}
\times\int
d^2x_\bot\exp\left[i(q_1+q_2+\Delta_\bot)x_\bot\right]K_0(\mu
\left|r_\bot\right|)
$$
$$
=\frac{1}{(2\pi)^2}\int d^2 q_1 d^2
q_2\times\frac{1}{(q_1^2+\mu^2)(q_2^2+\mu^2)\left[(q_1+q_2+\Delta_\bot)^2+\mu^2\right]}.\eqno(B.6)
$$

Apply the result that we obtained when calculating $I_3$ to this integral, we derive

$$
\int d^2q_1
\frac{1}{(q_1^2+\mu^2)\left[(q_1+q_2+\Delta_\bot)^2+\mu^2\right]}$$
$$=(-i\pi)\int_0^1 dx\frac{1}{\left[\mu^2+(q_2+\Delta_\bot)^2
x(1-x)\right]},
$$
so
$$
 I_4=\frac{1}{(2\pi)^2}(-i\pi)\int_0^1 dx\int d^2q_2\times \frac{1}{(q_2^2+\mu^2)\left[\mu^2+(q_2+\Delta_\bot)^2
x(1-x)\right]}\eqno(B.7)$$

From method of Feynman parameter integral, again, we obtain

$$I_4=-\frac{i}{(4\pi)}\int_0^1\frac{dx}{x(1-x)}\int_0^1dy\int
d^2q_2\times$$
$$\times\frac{1}{\left\{\left[(q_2+\Delta_\bot)^2+B\right]y+(q_2^2+\mu^2)(1-y)\right\}^2}$$
$$=-\frac{i}{(4\pi)}\int_0^1\frac{dx}{x(1-x)}\int_0^1dy\int
d^2q_2\frac{1}{\left(q_2^2+2q_2\Delta_\bot y+C\right)^2}$$
$$=-\frac{i}{(4\pi)}(-i\pi)\int_0^1\frac{dx}{x(1-x)}\int_0^1dy
\frac{1}{\left[C-(\Delta_\bot y)^2\right]}$$
$$=-\frac{1}{4}\int_0^1\frac{dx}{x(1-x)}\int_0^1dy
\frac{1}{\left[C-(\Delta_\bot y)^2\right]}.)$$

where

$$B=\frac{\mu^2}{x(1-x)},C=(\Delta_\bot^2+B)y+\mu^2(1-y)=\left[\frac{\mu^2}{x(1-x)}-t\right]y+\mu^2(1-y),\eqno(B.8)$$ then
$$I_4=-\frac{1}{4}\int_0^1\frac{dx}{x(1-x)}\int_0^1dy\times
\frac{1}{\left[\frac{\mu^2}{x(1-x)}-t\right]y+\mu^2(1-y)+ty^2}$$
$$=-\frac{1}{4}\int_0^1 dy\int_0^1 dx \frac{1}{-(1-y)(ty-\mu^2)x(1-x)+\mu^2}=-\frac{1}{4}\int_0^1 dy\int_0^1 dx \frac{1}{Dx^2-Dx+\mu^2}$$
$$=-\frac{1}{4}\int_0^1
\frac{dy}{D}\int_0^1\frac{dx}{x^2-x+\frac{\mu^2}{D}}=-\frac{1}{4}\int_0^1
\frac{dy}{D}\int_0^1\frac{dx}{(x-x_1)(x-x_2)}$$
$$
=-\frac{1}{4}\int_0^1\frac{dy}{D}\frac{1}{x_1-x_2}ln\left|\frac{(1-x_1)x_2}
{(1-x_2)x_1}\right|,\eqno(B.9)$$ here, $D=
-(1-y)(ty-\mu^2)=ty^2+(\mu^2-t)y+\mu^2$ and $x_1;x_2$ are roots of
equation $
x^2 -x+\frac{\mu^2}{D}=0 $.\\

Noting that: $x_1+x_2=1\Rightarrow 1-x_1=x_2;1-x_2=x_1$, and:
$$x_1-x_2=\sqrt{1-\frac{4\mu^2}{D}}\approx1-\frac{2\mu^2}{D},\eqno(B.10)$$
hence

$$ln\left|\frac{(1-x_1)x_2}{(1-x_2)x_1}\right|=ln\left|\frac{x_2^2}{x_1^2}\right|
=2ln\left|\frac{1-\sqrt{1-\frac{4\mu^2}{D}}}{1+\sqrt{1-\frac{4\mu^2}{D}}}\right|
\approx 2ln\left|\frac{\mu^2}{D-\mu^2}\right|.\eqno(B.11)$$

so that
$$I_4=-\frac{1}{2}\int_0^1 dy\frac{1}{D-2\mu^2}ln\left|\frac{\mu^2}{D-\mu^2}\right|$$
$$=-\frac{1}{2}\int_0^1 dy\frac{1}{(ty+\mu^2)(y-1)}ln\left|\frac{\mu^2}{y(ty+\mu^2-t)}\right|\equiv-\frac{1}{2}F_2(t)\eqno(B.12)$$

\end{document}